\documentclass[aps,prl,twocolumn,showpacs,preprintnumbers,superscriptaddress,amsmath]{revtex4}
\usepackage{txfonts}
\usepackage{amssymb}
\usepackage{graphicx}
\usepackage{dcolumn}
\usepackage{bm}

\begin{document}

\title{Interference of surface plasmon polaritions controlled by the phase of incident light}
\author{Xi-Feng Ren, Guo-Ping Guo\footnote{gpguo@ustc.edu.cn}, Yun-Feng Huang, Zhi-Wei Wang, Pei Zhang and Guang-Can Guo}
\address{Key Laboratory of Quantum Information, University of Science and Technology of China, Hefei
230026, People's Republic of China}

\begin{abstract}
Interference patterns of surface plasmon polaritons(SPPs) are
observed in the extraordinary optical transmission through
subwavelength holes in optically thick metal plate. It is found that
the phase of incident light can be transferred to SPPs. We can
control the destructive and constructive interference of SPPs by
modulating the relative phase between two incident beams. Using a
slightly displaced Mach-Zehnder interferometer, we also observe a
SPPs interference pattern composed of bright and dark stripes.

\end{abstract}

\maketitle

\newpage

Surface plasmon polariton(SPP) is a surface electromagnetic wave
coupling to the free electron oscillations in a metal. It can be
excited when a light wave strikes a metal film under appropriate
conditions. Such SPPs are involved in a wide range of
phenomena\cite{Barnes03,Ozbay06}, including nanoscale optical
waveguiding\cite{Bozhe06,Takah97,Taka04,Zia06}, perfect
lensing\cite{Pendry00}, extraordinary optical
transmission\cite{Ebbesen98}, subwavelength lithography\cite{Fang05}
and ultrahigh-sensitivity biosensing\cite{Leidberg83}. It has also
been experimentally proved that SPPs are also useful in the
investigation of quantum information\cite{alt,energy,ren062}. Since
surface plasmon-based photonics(plasmonics) has both the capacity of
photonics and the miniaturization of electronics, it may offer us a
solution to the size-compatibility problem\cite{Ozbay06}. To realize
the full potential technology of palsmonics, we need to construct a
general frame work to describe the propagating, diffraction and
interference of SPPs.

Interference of SPPs is first studied by the group of
Lezec\cite{Lezec06}. They have shown that light transmission through
a slit milled in an Ag film can be passively enhanced(suppressed) as
a result of constructive(destructive) interference with a SPP
launched by a groove nearby. Efficient unidirectional nanoslit
couplers for SPPs\cite{Lopez07} and all-optical modulation by
plasmonic excitation of CdSe quantum dots\cite{Paci07} are also
realized based on the interference of SPPs. A double-slit experiment
with SPPs is presented which reveals the analogue between SPPs
propagating along the surface of metallic structures and light
propagating in conventional dielectric components\cite{Zia07}. It is
also proved that SPPs can be excited with a focused laser beam at
normal incidence to a metal film without any protrusions and holes,
while the intensity distribution on the metal surface is partly
dominated by interference between counterpropagating
plasmons\cite{Bouh07}. In these works, the construction or
destruction of interference pattern is determined by the propagating
distance of SPPs on the metal surface. For example, in
\cite{Paci07}, the transmitted energy is varied with the distance
between the slit and the groove. Of course, many samples are needed
to give a full characterization of the interference pattern.

It is important for us to control the interference pattern of SPPs
on a given sample in the approach of the chip-based plasmonics. One
way is to change the wavelength of incident light, as in
\cite{Bozhe06}, where the transmission of the plasmonic
waveguide-ring resonator is varied with the light wavelength.
However, this method of varying wavelength may be not feasible in
the practical application. In this paper, we show that the phase of
the incident light can be transferred to the SPPs, which provides us
a potential method to modulate the interference pattern of SPPs
using the incident lights. We use two light beams to excite SPPs. By
controlling the relative phase of two beams, we can observe the
interference pattern of SPPs for serval periods. The transmission
can be tuned continuously from minimum to maximum with a ratio about
7, even when the power of the incident light is kept stable. To give
an intuitional illustration, we also take several pictures of the
interference pattern composed of bright and dark stripes using a
charge coupled device(CCD) camera behind a slightly displaced
Mach-Zehnder(MZ) interferometer. The present method may be valuable
in the future application of plasmonics due to the well-established
technology on the linear optical elements.

\begin{figure}[b]
\centering
\includegraphics[width=6.0cm]{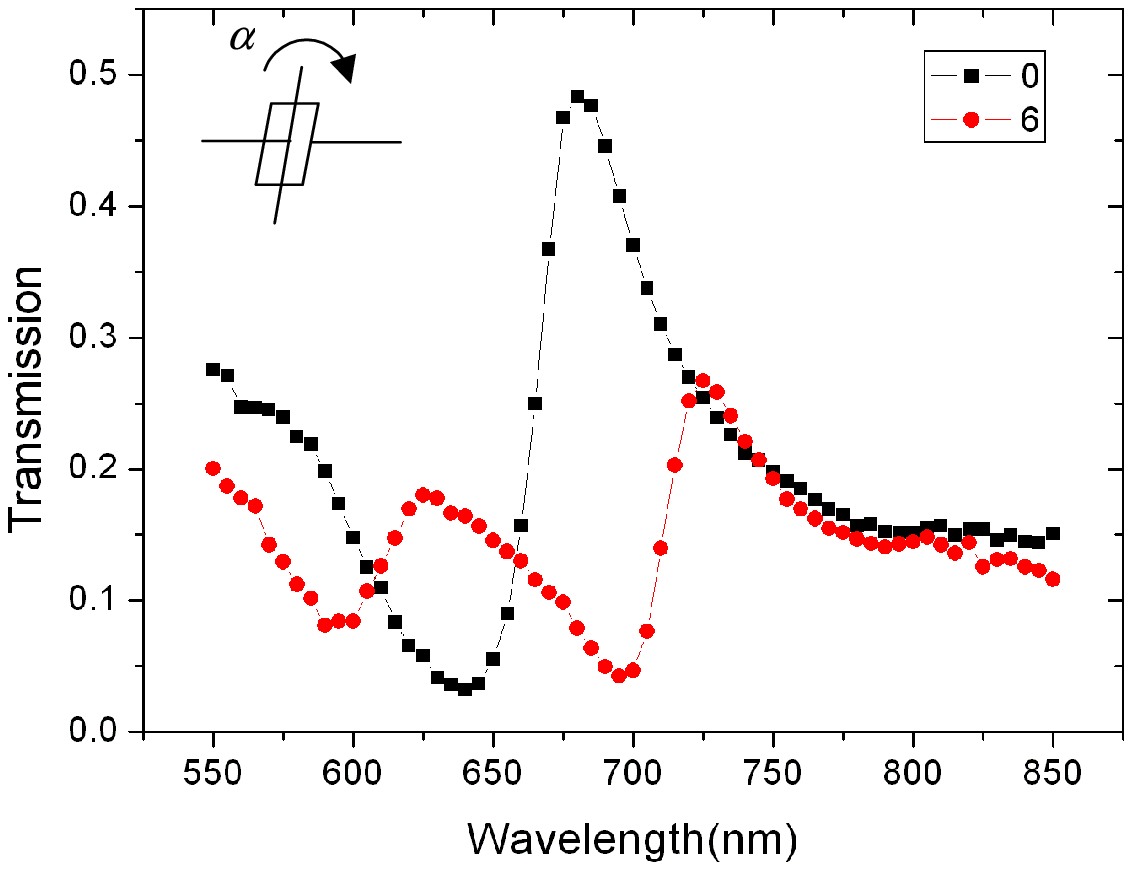}
\caption{(Color online)Hole array transmittance as a function of
wavelength for vertical polarized light with tilt angle
$\alpha=0\textordmasculine$(blue square dots) and
$6\textordmasculine$(red round dots). Inset is an illustration of
tilt direction.}
\end{figure}

The metal plate used in our experiment to excite SPPs is produced as
follows: after subsequently evaporating a $3$-$nm$ titanium bonding
layer and a $135$-$nm$ gold layer onto a $0.5$-$mm$-thick silica
glass substrate, a Electron Beam Lithography System (EBL, Raith 150
of Raith Co.) is used to produce cylindrical holes ($200nm$
diameter) arranged as a square lattice ($600nm$ period). The area of
the hole array is $300\mu m\times 300\mu m$. Transmission spectra of
the hole array are recorded by a Silicon avalanche photodiode (APD)
single photon detector coupled with a monochromator through a fiber.
White light from a stabilized tungsten-halogen source passes through
a single mode fiber and a polarizer (only vertical polarized light
can pass), then illuminates on the sample. The hole array is set
between two lenses with the focus of $35mm$, so that the light is
normally incident on the hole array with a cross sectional diameter
about $20\mu m$ and covers hundreds of holes. The light exiting from
the hole array is launched into the monochromator. The transmission
spectra are shown in Fig. 1. The black square dots are measured with
tilt angle $\alpha=0\textordmasculine$ and red round dots with
$\alpha=6\textordmasculine$. The reason for tilting the metal plate
will be explained below.

To control the interference pattern of SPPs by the incident lights,
firstly we need to prove that the phase of excited SPPs is
correlated with that of incident light. This is verified by using a
collinear polarization MZ interferometer as shown in Fig. 2. The
advantage of this kind of MZ interferometer is the stability against
the environment. We consider the case when the metal plate is
removed from the twin-lenses at first. White light from a stabilized
tungsten-halogen source passes through single mode fiber and $4 nm$
filter (center wavelength 702 nm) to provide photons at the
wavelength of 702nm. Only vertical polarized photons can transmit
the first polarization beam splitter(PBS). When the first half wave
plate(HWP) is rotated to $-22.5\textordmasculine $, the photons in
the state $\left| V\right\rangle $ will be changed into the state
$(\left| H\right\rangle +\left| V\right\rangle )/\sqrt{2}$. Then
these photons go through the birefringent crystal(BC), where they
get a phase difference $\Delta \varphi $ between horizontal and
vertical polarization modes. The state is thus in the form of
$(\left| H\right\rangle +e^{i\Delta \varphi }\left| V\right\rangle
)/\sqrt{2}$. After the second HWP($+22.5\textordmasculine $), the
state is transformed into $1/2((1+e^{i\Delta \varphi })\left|
H\right\rangle+(1-e^{i\Delta \varphi })\left| V\right\rangle)$. Then
the photons are separated by the second PBS which also only permits
the transmission of $V$ polarized photons. Since there is a phase
difference between the two parts of transmitted $V$ polarized
photons, the counts can be varied with this phase difference due to
the interference effect. The experiment results are shown in Fig. 3a
(Black square dots), which fits the theoretical interference pattern
nicely in the $\left| V\right\rangle $ basis
\begin{equation}
R_{V}=(\sin (\Delta \varphi/2))^2.
\end{equation}
Now we put the metal plate with hole array between the twin lenses.
The light irradiates the metal plate normally and the transmission
efficiency is measured. In this case, photons are first transformed
into SPPs and then back to photons\cite{Ebbesen98}. Fig. 3a(Red
round dots) are the experimental results, which also fits the
theoretical calculation nicely. This gives the evidence that the
phase of the input light can be transferred to the SPPs.

\begin{figure}[b]
\centering
\includegraphics[width=8.0cm]{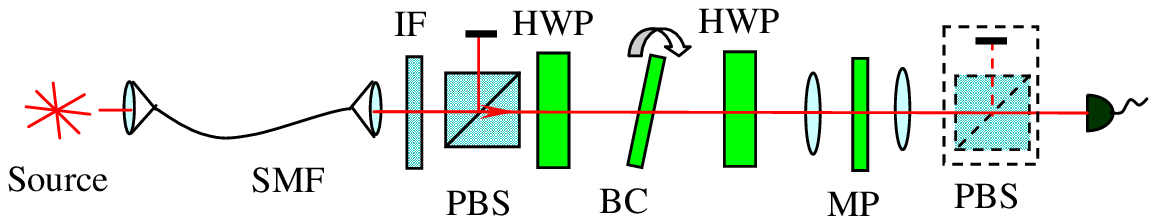}
\caption{(Color online)Experimental setup to observe the
interference pattern of SPPs for $702nm$ wavelength light. The
collinear polarization MZ interferometer is composed of two PBS, two
HWPs and a birefringent crystal. The birefringent crystal is tilted
to generate a phase difference $\Delta \varphi $ between horizontal
and vertical polarization modes. The metal plate is set between two
lenses with the focus of $35mm$ and the second PBS is removed in
some cases.}
\includegraphics[width=8.0cm]{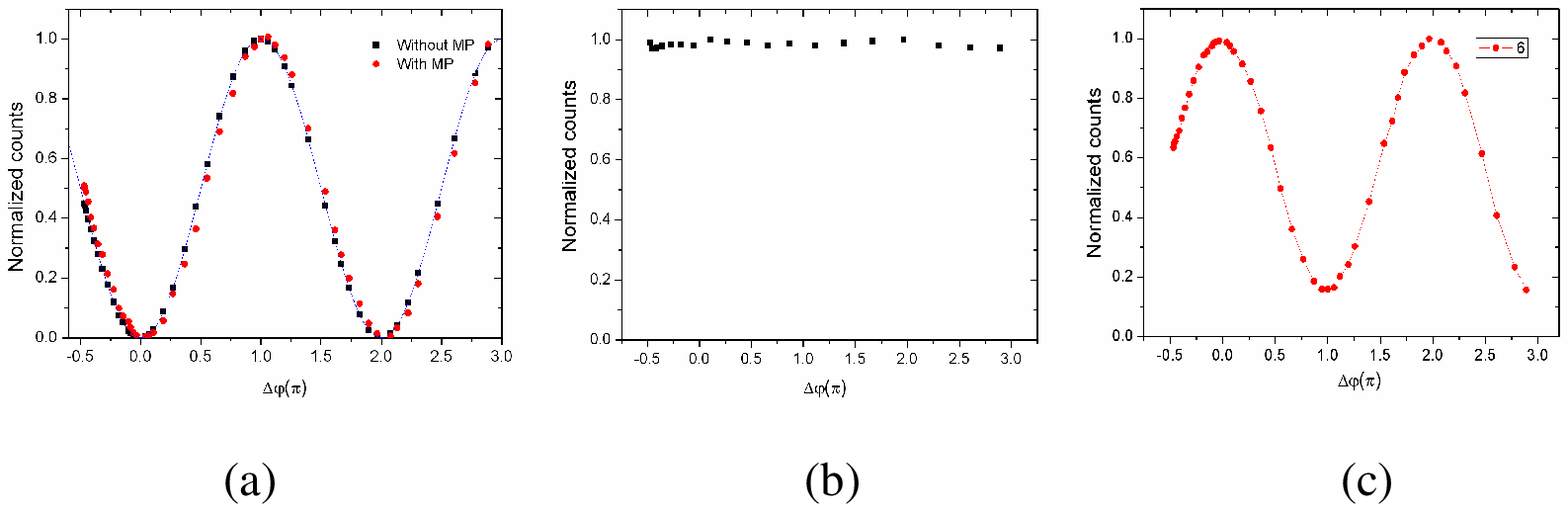}
\caption{(Color online)Interference pattern for light with $702nm$
wavelength. (a) The second PBS is placed in the experimental setup.
Black square dots are normalized counts without the metal plate,
while red round dots with the metal plate. They both fit the
theoretical calculation(the line) well, which sustain the conclusion
that the phase of the illuminating light can be transferred to the
SPPs. (b) The counts are kept constant when we take the second PBS
out and the metal plate away. (c) An interference pattern is
observed when the metal plate is placed between the two lenses at
the tilted angle of $6\textordmasculine$, even without the second
PBS. The interference of SPPs can be fully controlled by the phase
difference $\Delta \varphi$.}
\end{figure}

\begin{figure}[b]
\centering
\includegraphics[width=8.0cm]{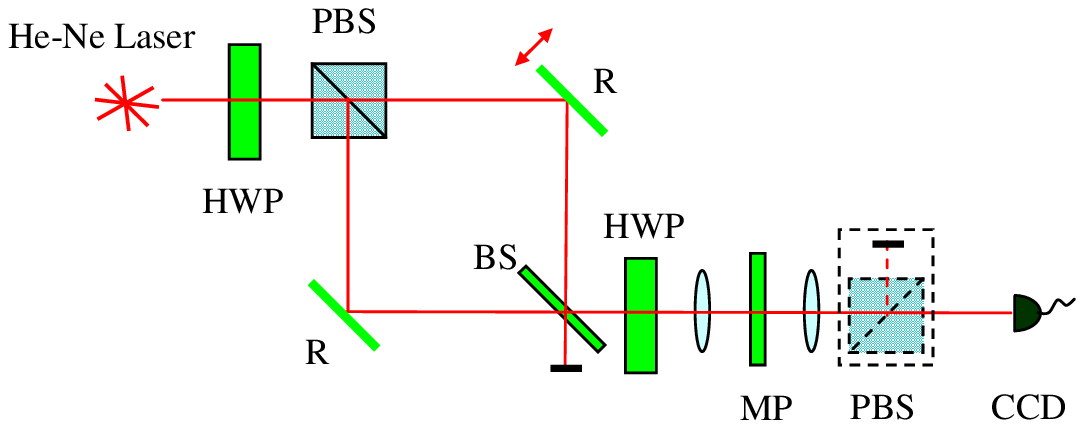}
\caption{(Color online)Experimental setup to observe the
interference pattern of SPPs for $632.8nm$ wavelength light. The
path difference of MZ interferometer gives the phase difference
$\Delta\varphi$. A reflector is displaced slightly to observe the
interference pattern of bright and dark stripes.}
\includegraphics[width=8.0cm]{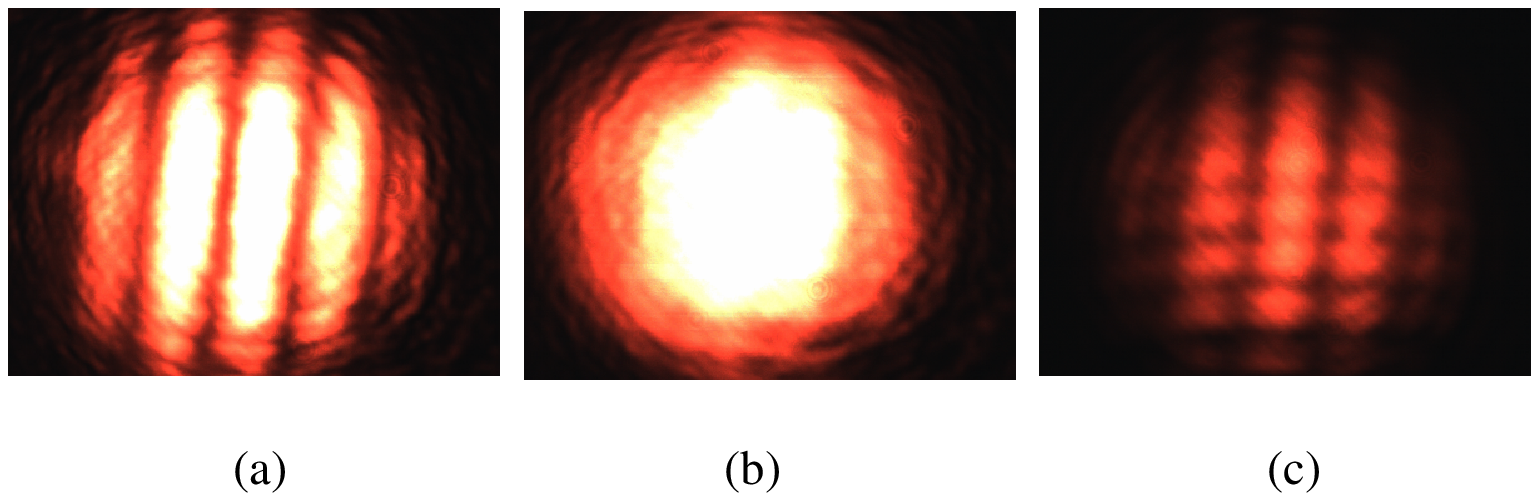}
\caption{(Color online)Energy distribution recorded by the CCD. (a)
General interference pattern of a displaced MZ interferometer with
the second PBS.(b) Energy distribution when the second PBS is
removed.  The metal plate is moved out for (a) and (b). (c)
Interference pattern of SPPs is observed even when the second PBS is
removed. The metal plate is tilted $6\textordmasculine$ to excite
SPPs.}
\end{figure}

However, this curve may result from the interference of the
transmitted photons on the second PBS, not the SPPs on the metal
surface. It is necessary for us to do a further investigation. We
remove the second PBS from the setup to avoid this probability. It
is observed that the counts are kept constant and there is no
interference phenomenon as shown in Fig. 3b when the metal plate is
moved away. The reason is that all the photons are detected in this
case. Then the metal plate is placed between the two lenses at the
tilted angle of $\alpha=6\textordmasculine$. In this case, due to
the removal of SPPs degeneracy, the transmission spectra is changed
for the vertically polarized photons\cite{Ebbesen98,Genet,Ren071}.
The transmission peak at 680nm is divided into two peaks at the
wavelength of 625nm and 725nm respectively. It is noted that the
transmission spectra for horizontal polarized photons is not
influenced. The transmission efficiency of 702nm light with vertical
polarization is only $0.051\%$ at the tilted angle of
$6\textordmasculine$, which comes from the direct transmission from
the nano-scale holes. While for horizontal polarized light, the
transmission is about $0.354\%$ due to the SPPs assisted
transmission process\cite{Ebbesen98}. Thus most of the counts are
coming from the horizontal polarized photons. A clear interference
pattern is observed in this situation as shown in Fig. 3c. This
phenomenon must come from the interference of SPPs excited on the
metal surface by the horizontal polarized light. Of course, the
pattern is contrary to the previous cases as shown in Fig. 3a. The
ratio between maximum transmission and minimum transmission is about
7, corresponding to the ratio of transmission efficiency between the
horizontal and vertical polarized lights, which is determined by
titled angle $\alpha$ of the metal plate . So we can draw the
conclusion that the interference of SPPs can be expediently
controlled by the phase of illuminated light with linear optical
elements.

We also get an intuitional illustration of the interference of SPPs.
A path MZ interferometer is used as shown in Fig. 4. The essence of
this MZ interferometer is equal to the previous collinear
polarization interferometer, while the phase $\Delta\varphi$ comes
from the path difference. In this interferometer, the position of
two light beams can be tuned slightly by the reflector(R)(see Fig.
4). The collinear of the two light beams after the beam splitter(BS)
is slightly destroyed in the horizontal direction by displacing the
reflector. An interference pattern with bright and dark stripes in
the vertical direction is recorded by a CCD when the second PBS is
placed in the setup(Fig. 5a). If we take the second PBS away, there
is no interference pattern(Fig. 5b). When we put the metal plate in
at the tilted angle of $6\textordmasculine$, a similar image of
bright and dark stripes appears as shown in Fig. 5c, which is a
vivid picture of interference of SPPs. The three pictures correspond
to the cases of Fig. 3a(Black square dots), Fig. 3b and Fig. 3c
respectively. Since the wavelength of incident light is $632.8 nm$,
the SPPs are excited by the vertical polarized light, which is
different from the case of $702 nm$.

In conclusion, interference of SPPs is observed by controlling the
phase of incident light. Due to the well-established technology on
the linear optical elements, we can modulate the interference
pattern of SPPs from destructive to constructive continuously. Our
method may be useful in the future investigation of plasmonics.

This work was funded by the National Fundamental Research
Program(No. 2006CB921900), National Natural Science Foundation of
China (No. 10604052), Chinese Academy of Sciences International
Partnership Project, the Program of the Education Department of
Anhui Province (Grant No.2006kj074A). Xi-Feng Ren also thanks for
the China Postdoctoral Science Foundation (20060400205) and the K.
C. Wong Education Foundation, Hong Kong.

\end{document}